\documentclass[preprint,showpacs,amsmath]{revtex4-1}
\def\qed{\leavevmode\unskip\penalty9999 \hbox{}\nobreak\hfill
     \quad\hbox{\leavevmode  \hbox to.77778em{%
               \hfil\vrule   \vbox to.675em%
               {\hrule width.6em\vfil\hrule}\vrule\hfil}}
     \par\vskip3pt}
\usepackage{epsfig}
\newtheorem{theorem}{Theorem}

\begin{document}

\begin{center}
{Quantum correlations induced by local von Neumann measurement}
\end{center}

\begin{center}
{Ming-Jing Zhao$^{1}$},
{Ting-Gui Zhang$^{1}$},
{Zong-Guo Li$^{2}$},
{Xianqing Li-Jost$^{1}$},
{Shao-Ming Fei$^{1,3}$}, and
{De-Shou Zhong$^{4}$}

\small {$^1$Max-Planck-Institute for Mathematics in the Sciences, 
Leipzig, 04103, Germany\\
{$^{2}$College of Science, Tianjin University of
Technology, Tianjin, 300191, China}\\
$^3$School of Mathematical Sciences, Capital Normal
University, Beijing 100048, China\\
$^4$Center of Mathematics, China Youth University for Political Sciences, Beijing, 100089, China}

\end{center}

{\bf Abstract}
We study the total quantum correlation, semiquantum correlation and joint quantum correlation induced by local von Neumann measurement in bipartite system. We analyze the properties of these quantum correlations and obtain analytical formula for pure states. The experiment witness for these quantum correlations is further provided and the significance of these quantum  correlations is discussed in the context of local distinguishability of quantum states.



\section{Introduction}
Quantum systems are correlated in a way that is inaccessible to classical ones. Furthermore, the correlations have advantages for quantum computing and information processing.
In recent years, many attention therefore have been paid to quantify quantum correlation and different measures have been proposed from different aspects.
In bipartite system, discord \cite{H. Ollivier,L. Henderson} is first introduced to quantify quantum correlation, which is defined as the difference between two quantum analogues of
the classical mutual information.
It has been shown that almost all quantum states have nonvanishing discord \cite{A. Ferraro}. Based on this, Ref. \cite{B. Daki} puts forward a geometric way of quantifying discord and obtains a closed form of expression for two-qubit state.
In Refs. \cite{S. Luo2008,S. Luo2010}, they investigate the quantum correlation
induced by local von Neumann measurement and reveal its equivalence with
the geometry of discord. In multipartite systems, a unified view of correlations has been discussed using relative entropy and
square norm as the distance respectively \cite{K. Modi,B. Bellomo}. Then Ref. \cite{G. L. Giorgi} defines the genuine correlation as the amount of correlation that can not be accounted for considering any of the possible subsystems. Recently, postulates for measures of quantum correlations have been proposed \cite{A. Brodutch,C. H. Bennett2011}, but the quantum correlation is still far from being understood.

Since quantum states can be divided into classical state, semiquantum state and truly quantum state,
quantum correlations are classified into total quantum correlation, semiquantum correlation and joint quantum correlation in terms of local von Neumann measurement \cite{ M. Gessner, S. Luo2008,S. Luo2010}. We study the properties of these quantum correlations and calculate the analytical formula of
these quantum correlations for pure states. It is shown that these quantities coincide for pure state and are proportional to the squared
concurrence. Furthermore,
we provide a witness for experimental detection of these quantum correlations by employing the strategy in Ref. \cite{M. Gessner}. The result is finally applied to local distinguishability of quantum states, showing that the nonexisting joint quantum correlation is the necessary condition for distinguishing two-qubit separable and orthogonal pure states locally.

\section{Quantum correlations induced by local von Neumann measurement}

Let $H_m$ and $H_n$ denote $m$ and $n$ dimensional complex Hilbert spaces,
with $\{\vert i\rangle\}_{i=1}^m$ and $\{\vert j\rangle\}_{j=1}^n$ the orthonormal basis for $H_m$ and $H_n$ respectively.
Let $\rho$ be a density matrix defined on $H_m\otimes H_n$.
We call $\rho$  a classical correlated (C-C) state if $\rho=\sum_{ij} p_{ij} |ij\rangle \langle ij|$, $0\leq p_{ij} \leq 1$, $\sum_{ij} p_{ij}=1$.  $\rho$ is called classical-quantum (C-Q) correlated if
$\rho=\sum_i p_i |i\rangle \langle i|\otimes \rho_i$, with $\rho_i$ the density matrices on $H_n$, $0\leq p_{i} \leq 1$, $\sum_{i} p_{i}=1$. Analogously,
$\rho=\sum_j p_j \rho_j \otimes |j\rangle \langle j|$ is said to be a quantum-classical (Q-C) correlated state.
For short, the C-Q and Q-C states are called semiquantum ones which can be of both quantum and classical correlations \cite{L. Henderson,K. Modi}.

Let $\Phi_1=\{\pi^{(1)}_u\}$ and $\Phi_2=\{\pi^{(2)}_v\}$ stand for the local von Neumann measurements acting unilaterally on the first and second subsystems respectively,
\begin{equation}
\begin{array}{rcl}
\Phi_1(\rho)=\sum_{u} \pi^{(1)}_u\otimes I \rho \pi^{(1)}_u \otimes I, \\
\Phi_2(\rho)=\sum_{v} I\otimes\pi^{(2)}_v  \rho I \otimes \pi^{(2)}_v.
\end{array}
\end{equation}
Let $\Phi_{12}\equiv\Phi_1\circ \Phi_2$ be the local von Neumman measurements acting bilaterally on $\rho$,
\begin{eqnarray}
\Phi_{12}(\rho)=\sum_{u,v} \pi^{(1)}_u \otimes \pi^{(2)}_v \rho \pi^{(1)}_u\otimes \pi^{(2)}_v.
\end{eqnarray}
Generally, $\Phi_{i}(\rho)$ is semiquantum state and $\Phi_{12}(\rho)$ is classical state. The quantum correlation is then defined as the change of the quantum state $\rho$ induced by the local von Neumann measurement.

First, for arbitrary quantum state $\rho$, the distance minimized
under all local von Neumann measurements $\Phi_{i}$,
\begin{eqnarray}
Q_{i}(\rho)&\equiv& \min_{\Phi_{i}}Q_{i}(\Phi_{i},\rho)\nonumber\\
&=&\min_{\Phi_{i}}||\rho-\Phi_{i}(\rho)||^2\nonumber\\
&=&tr(\rho^2)-\max_{\Phi_{i}}tr[(\Phi_{i}(\rho))^2],
\end{eqnarray}
is defined as quantum correlation with respect to the $i$-th part, $i=1,2$ \cite{S. Luo2010}. Here the Hilbert-Schmidt norm $||A||=\sqrt{tr(A^\dagger A)}$ has been used as the measure of distance. We call $Q_{i}(\rho)$ semiquantum correlation typically compared with the total quantum correlation, $i=1,2$. $Q_{1}(\rho)=0$ (resp. $Q_{2}(\rho)=0$) if and only if $\rho$ is a C-Q (resp. Q-C) state.

Based on these, we define
the total quantum correlation $Q_{12}(\rho)$ as the minimized distance
\begin{eqnarray}
Q_{12}(\rho)&\equiv&\min_{\Phi_{12}}Q_{12}(\Phi_{12},\rho)\nonumber\\
&=&\min_{\Phi_{12}}||\rho-\Phi_{12}(\rho)||^2\nonumber\\
&=&tr(\rho^2)-\max_{\Phi_{12}}tr[(\Phi_{12}(\rho))^2].
\end{eqnarray}
$Q_{12}(\rho)=0$ if and only if $\rho$ is a C-C state.

The sum of the semiquantum correlations is generally larger than the total quantum correlation. We call this discrepancy the joint quantum correlation,
\begin{eqnarray}
\delta(\rho)\equiv Q_{1}(\rho)+Q_{2}(\rho)-Q_{12}(\rho).
\end{eqnarray}

The geometric picture of these quantum correlations is clear and illustrated in Fig. 1. In fact the set of C-Q states,
Q-C states and C-C states are not convex. The overlap between the set of C-Q states and the set of Q-C
states are the set of C-C states. For any given quantum state $\rho$, $Q_1(\rho)$ (resp. $Q_2(\rho)$)
is the minimum distance between
$\rho$ and the set $\{\Phi_1(\rho)\}$ (resp. $\{\Phi_2(\rho)$\}) under all local von Neumann measurements
on the first (resp. second) subsystem,
and
$Q_{12}(\rho)$ is the minimum distance between
$\rho$ and the set $\{\Phi_{12}(\rho)\}$ under all local von Neumann measurements on both subsystems.

\begin{center}
\begin{figure}[!h]\label{geo}
\resizebox{8.5cm}{!}{\includegraphics{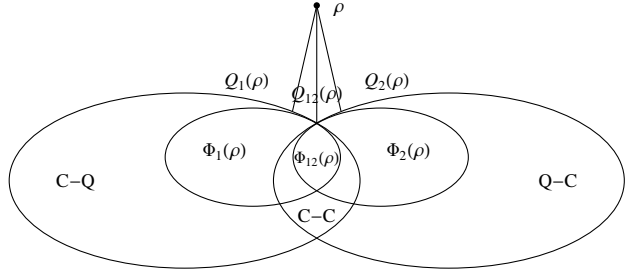}}
\caption{The geometric picture of the quantum correlations $Q_1(\rho)$, $Q_2(\rho)$ and $Q_{12}(\rho)$.}
\end{figure}
\end{center}

In Ref. \cite{S. Luo2010}, it shows $Q_1({\rho})$ is equal to the geometric measure of discord which is defined as $D_1({\rho})=\min_{\sigma\in\Omega_1}||\rho-\sigma||^2$ with $\Omega_1$ the set of all C-Q states \cite{B. Daki}, i.e. $Q_1({\rho})=D_1({\rho})$. Thus for any quantum state $\rho$, the nearest C-Q state induced by local von Neumann measurement on the first subsystem is just the nearest one comparing all the C-Q states. Furthermore, we can get a general conclusion that, for any quantum state, the nearest semiquantum state or classical state induced by local von Neumann measurement is the nearest one comparing all the semiquantum states or classical states.

The total quantum correlation $Q_{12}(\rho)$, semiquantum correlations $Q_{i}(\rho)$ ($i=1,2$) and the joint quantum correlation $\delta(\rho)$ have the following properties.

(i) $0\leq \delta(\rho)\leq Q_{i}(\rho)\leq Q_{12}(\rho)<1$, for $i=1,2$. 

First, $Q_{12}(\rho)<1$ is obvious by definition. Second, for any quantum state $\rho$, assume
$\Phi^{\prime\prime}_{1}$ and $\Phi^{\prime\prime}_{2}$ are the optimal local von Neumann measurements acting on the first and second subsystems respectively such that
they reach the minimum of total quantum correlation,
i.e. $Q_{12}(\rho)=||\rho-\Phi^{\prime\prime}_{12}(\rho)||^2$, $\Phi^{\prime\prime}_{12}=\Phi^{\prime\prime}_{1}\circ \Phi^{\prime\prime}_{2}$, we have
\begin{eqnarray*}
Q_{12}(\rho)- Q_{i}(\rho)
\geq ||\rho-\Phi^{\prime\prime}_{12}(\rho)||^2-||\rho-\Phi^{\prime\prime}_i(\rho)||^2
=tr\{[\Phi^{\prime\prime}_i(\rho)-\Phi^{\prime\prime}_{12}(\rho)]^2\}\geq 0,
\end{eqnarray*}
for $i=1,2$. Third, since 
\begin{eqnarray*}
Q_1(\rho)-\delta(\rho)=Q_{12}(\rho)- Q_{2}(\rho)\geq 0,\\
Q_2(\rho)-\delta(\rho)=Q_{12}(\rho)- Q_{1}(\rho)\geq 0,
\end{eqnarray*}
so $Q_i(\rho)\geq \delta(\rho)$ for $i=1,2$. Fourth,
to show $\delta(\rho)\geq 0$, we
suppose $\Phi^\prime_{i}$ is the optimal local von Neumann measurement acting on the $i$-th subsystem that reaches the
minimum of the semiquantum correlation for $\rho$, i.e.
$Q_{i}(\rho)=||\rho-\Phi^\prime_{i}(\rho)||^2$ for $i=1,2$.
Then
\begin{eqnarray*}
\delta(\rho)\geq tr\{[\rho-\Phi^\prime_1(\rho)-\Phi^\prime_2(\rho)+\Phi^\prime_{12}(\rho)]^2\}\geq 0.
\end{eqnarray*}

(ii) If $Q_{i}(\rho)=0$ or $Q_{12}(\rho)=0$, $i=1,2$, then $\rho$ has no joint quantum correlation, $\delta(\rho)=0$.

(iii) All these quantum correlations are invariant under local unitary operations.

According to the definitions and properties of these quantum correlations, we have the following theorem whose format is analogous to the one in Ref. \cite{S. Luo2012}.
\begin{theorem}\label{th}
Let $|\psi\rangle$ be any bipartite pure state with Schmidt decomposition $|\psi\rangle=\sum_{i} \lambda_i |ii\rangle$, $0\leq \lambda_{i}\leq 1$ and $\sum_i \lambda_i^2=1$. The total quantum correlation, semiquantum correlation and joint quantum correlation coincide, i.e. $Q_{12}(|\psi\rangle)=Q_{1}(|\psi\rangle)=Q_{2}(|\psi\rangle)=\delta(|\psi\rangle)=1-\sum_i \lambda_i^4$.
\end{theorem}

Proof. Let $\Phi_1=\{\pi_j^{(1)}\}=\{|\psi_j\rangle \langle\psi_j|\}$ and $\Phi_2=\{\pi_j^{(2)}\}=\{|\phi_j\rangle \langle\phi_j|\}$ be two arbitrary local von Neumann measurements acting on the first and second subsystem respectively, with
\begin{equation}\label{constraint th}
\begin{array}{rcl}
&&|\psi_j\rangle=\sum_i a_{ij}|i\rangle,\  \sum_j |\psi_j\rangle \langle\psi_j|=I, \ \langle\psi_j|\psi_{j^\prime}\rangle=\delta_{jj^\prime},\\[1mm]
&&|\phi_j\rangle=\sum_i b_{ij}|i\rangle, \ \sum_j |\phi_j\rangle \langle\phi_j|=I, \ \langle\phi_j|\phi_{j^\prime}\rangle=\delta_{jj^\prime}.
\end{array}
\end{equation}
After the local von Neumann measurement $\Phi_1$ acting on the first subsystem, quantum state $|\psi\rangle$ becomes
\begin{eqnarray*}
\Phi_1(|\psi\rangle)
=\sum_j |\psi_j\rangle\langle\psi_j|\otimes (\sum_i a_{ij}^*\lambda_i |i\rangle)(\sum_i a_{ij} \lambda_i \langle i|),
\end{eqnarray*}
and $tr[(\Phi_1(|\psi\rangle))^2] =\sum_j (\sum_i \lambda_i^2 |a_{ij}|^2)^2$. In order to find the minimum of $Q_1(|\psi\rangle)$ and the nearest C-Q state $\Phi_1(|\psi\rangle)$ to $|\psi\rangle$, we need to find the maximum of $tr[(\Phi_1 (|\psi\rangle))^2]$ under the constraints in Eq. (\ref{constraint th}). Consider the multivariable function
\begin{eqnarray*}
f(x)=\sum_j (\sum_i \lambda_i^2 x_{ij})^2
\end{eqnarray*}
with the restrictions
\begin{eqnarray*}
0\leq x_{ij}\leq 1;~~
\sum_{j}x_{ij}=1, \, \forall i;~~
\sum_{i}x_{ij}=1, \, \forall j,
\end{eqnarray*}
it can be verified that the function $f(x)$ is convex with respect to the variables $x_{ij}$, as its Hessian matrix is nonnegative. Therefore its maximum is attained at the boundary, $x_{ij}=0,1$, $\forall i,j$. Hence we further derive that the maximum of $f(x)$ is $\sum_i \lambda_i^4$ which is attained when the matrix $X=(x_{ij})$ is a permutation matrix. Therefore we have $\max_{\Phi_1} tr[(\Phi_1(|\psi\rangle))^2]=\sum_i \lambda_i^4$ if we choose $|\psi_j\rangle=|j\rangle$, $\forall j$. This implies $Q_{1}(|\psi\rangle)=1-\sum_i \lambda_i^4$ and $\rho^\prime=\sum_i \lambda_i^2 |i\rangle\langle i|\otimes |i\rangle\langle i|$ is one of the nearest C-Q states to $|\psi\rangle$. Similarly, one can get the result for $Q_{2}(|\psi\rangle)$.

For the total quantum correlation $Q_{12}(|\psi\rangle)$, since
$Q_1(|\psi\rangle)\leq Q_{12}(|\psi\rangle)$, we get 
$$\max_{\Phi_{12}}tr [(\Phi_{12}(|\psi\rangle))^2]\leq
\max_{\Phi_1}tr [(\Phi_1(|\psi\rangle))^2]
=\sum_i \lambda_i^4$$ 
and the inequality becomes equality when $|\psi_j\rangle=|\phi_j\rangle=|j\rangle$, $\forall j$.
As a result $Q_{12}(|\psi\rangle)=1-\sum_i \lambda_i^4$ and $\rho^\prime$ is also the nearest C-C state to $|\psi\rangle$. Finally, it is direct to get $\delta(|\psi\rangle)=1-\sum_i \lambda_i^4$ by definition.
\qed

Here it is worth to mention that for any bipartite pure state $|\psi\rangle$,
its quantum correlations are
proportional to the squared concurrence \cite{concurrence,fei,Rungta},
$Q_{1}(|\psi\rangle)=Q_{2}(|\psi\rangle)=Q_{12}(|\psi\rangle)=\delta(|\psi\rangle)=C^2(|\psi\rangle)/2$.
Hence the quantum correlations for pure states are just the quantum entanglement \cite{L. Henderson}.

In addition, if one restricts the local von Neumann measurements to the ones that keep the marginal states invariant,
one can similarly define total quantum correlation, semiquantum correlation and joint quantum correlation. In this way, the  properties (i)-(iii) and Theorem \ref{th} still hold true.

For general bipartite mixed state $\rho$, the quantum correlation can be estimated in view of Ref. \cite{S. Luo2010} as follows.
\begin{theorem}
For any mixed state $\rho\in H_m\otimes H_n$ ($m\leq n$),
$\rho=\sum_{i=1}^{m^2}\sum_{j=1}^{n^2} c_{ij} X_{i}\otimes Y_{j}$, where $X_1=I_m$, $Y_1=I_n$, $X_{i}$ and $Y_i$ are the generators of $SU(m)$ and $SU(n)$ respectively, $i=2,\cdots, m^2$, $j=2,\cdots, n^2$,
we have
\begin{equation*}
\begin{array}{rcl}
Q_1{(\rho)}&=&tr(CC^T)-\max_A tr(ACC^TA^T),\\
Q_2{(\rho)}&=&tr(CC^T)-\max_B tr(BC^TCB^T),\\
Q_{12}(\rho)&=&tr(CC^T)-\max_{A,B} tr(ACB^TBC^TA^T),
\end{array}
\end{equation*}
where $C=(c_{ij})$, $A=(a_{ki})$ such that $a_{ki}=tr (|k\rangle\langle k| X_i)$ for $k=1,\cdots, m$ and $i=1, \cdots, m^2$, $B=(b_{lj})$ such that $b_{lj}=tr (|l\rangle\langle l| Y_i)$ for $k=1,\cdots, n$ and $i=1, \cdots, n^2$ with $\{|k\rangle\}$
and $\{|l\rangle\}$ any orthonormal basis for the first and second subsystems respectively.
Especially, $Q_1(\rho),\ Q_{12}(\rho)\geq \sum_{i=m+1}^{m^2}\lambda_i$, and
$Q_2(\rho)\geq \sum_{i=n+1}^{n^2}\lambda_i$,
where $\lambda_i$ are the eigenvalues of $CC^T$ listed in decreasing order.
\end{theorem}

In fact, these quantum correlations can be calculated exactly for some special quantum states. For the isotropic states \cite{M. Horodecki1999}:
\begin{eqnarray*}
\rho_1(f_1)
&=&\frac{1-f_1}{n^2-1} I_n +  \frac{n^2f_1-1}{n^2-1} |\psi^+ \rangle
\langle \psi^+|,
\end{eqnarray*}
with $f_1=\langle \psi^+| \rho_1(f_1) |\psi^+ \rangle$ satisfying
$0\leq f_1 \leq 1$, $|\psi^+ \rangle=\frac{1}{\sqrt{n}}\sum_{i=0}^{n-1}|ii\rangle$,
we have the semiquantum correlation and total quantum correlation are $\frac{(n^2f_1-1)^2}{n(n+1)^2(n-1)}$. While, for the
Werner states \cite{werner}:
\begin{eqnarray*}
\rho_2(f_2) = \frac{n-f_2}{n^3-n} I_n + \frac{nf_2-1}{n^3-n}{V},
\end{eqnarray*}
where ${V}=\sum_{i,j=0}^{n-1} |ij\rangle \langle ji|$ and $f_2=\langle
\psi^+| \rho_2(f_2) |\psi^+ \rangle$, $-1\leq f_2 \leq 1$, we get the semiquantum correlation and total quantum correlation are $\frac{(nf_2-1)^2}{n(n+1)^2(n-1)}$. So for these two classes of states, the joint quantum correlation is the same as the semiquantum correlation and total quantum correlation.
Furthermore they have no quantum correlation if and only if they are the maximally mixed state.
Thus almost all isotropic states and Werner states
have nonzero quantum correlations.

For quantum correlation detection, notice that $tr(\rho^2)=tr[(P^+-P^-)\rho^{\otimes 2}]=1-2tr(P^-\rho^{\otimes 2})$, where $P^\pm$ are the projectors on the symmetric and antisymmetric subspaces respectively \cite{M. Hendrych}, so this provides an experimental way of measuring quantum correlations $Q_{i}(\rho)$ and  $Q_{12}(\rho)$, $i=1,2$.
For any given local von Neumann measurements $\Phi_1$ and $\Phi_2$,
the experimentally accessible witness for
$Q_{1}(\Phi_{1},\rho)$, $Q_{2}(\Phi_{2},\rho)$ and
$\delta(\Phi_1,\Phi_2,\rho)\equiv Q_{1}(\Phi_{1},\rho)+Q_{2}(\Phi_{2},\rho)-Q_{12}(\Phi_{1},\Phi_{2},\rho)$
can also be constructed by
using the strategy proposed in Ref. \cite{M. Gessner}.
Consider now the dynamics of the total system given by some unitary operator $U$. If $\rho$ has nonzero total quantum correlation (resp. semiquantum correlation), namely, $\rho$ and $\Phi_{12}(\rho)$ (resp. $\Phi_{i}(\rho)$) are not identical, then it has
\begin{equation}\label{correlation witness-1}
\langle||tr_B\{U(\rho-\Phi_{12(i)}(\rho))U^\dagger\}||^2\rangle=f(m,n)Q_{12(i)}(\rho),
\end{equation}
for $i=1,2$, and
\begin{eqnarray}\label{correlation witness-2}
\langle||tr_B\{U(\rho-\Phi_1(\rho)-\Phi_2(\rho)+\Phi_{12}(\rho))U^\dagger\}||^2\rangle
=f(m,n)\delta(\rho),
\end{eqnarray}
where $f(m,n)=\frac{m^2n-n}{m^2n^2-1}$, the expectation value of the function $F(U)$ is $\langle F(U)\rangle \equiv \int d\mu(U) F(U)$, $d\mu(U)$ is the probability measure on the unitary group. It shows the expectation value of the distance between the quantum state and the corresponding classical states (resp. semiquantum states) is proportional to the total quantum correlation (resp. semiquantum correlation), with the prefactor depending only on the dimensions of the subsystems. The expectation value of this witness in Eq. (\ref{correlation witness-1}) or Eq. (\ref{correlation witness-2}) with respect to randomly drawn unitaries is nonzero if and only if the initial state contains total quantum correlation (resp. semiquantum correlation) or joint quantum correlation respectively.

As an application of the joint quantum correlation we consider the problem of local distinguishability of quantum states. A set of bipartite pure states is exactly locally distinguishable if there is some sequence of local operations and classical communications (LOCC) that determines with certainty which state it is. The Bell states present a simple example of an orthogonal set that is not locally distinguishable \cite{S. Ghosh, J. Walgate}.
In Ref. \cite{J. Walgate} it has been proved in two-qubit system that if two separable and orthogonal states $|\psi_1\rangle$ and $|\psi_2\rangle$ are locally distinguished, then they must be of the form $\{|\psi_1\rangle=\frac{1}{\sqrt{2}}(|00\rangle+|10\rangle),\  |\psi_2\rangle=\frac{1}{\sqrt{2}}(|01\rangle+|11\rangle)\}$ or $\{|\psi_1\rangle=\frac{1}{\sqrt{2}}(|00\rangle+|01\rangle),\  |\psi_2\rangle=\frac{1}{\sqrt{2}}(|10\rangle+|11\rangle)\}$. Three separable and orthogonal states are locally distinguished if and only if they have the form, $\{|\psi_1\rangle=\frac{1}{\sqrt{2}}(|00\rangle+|10\rangle),\ |\psi_2\rangle=|01\rangle,\ |\psi_3\rangle=|11\rangle\}$ or $\{|\psi_1\rangle=\frac{1}{\sqrt{2}}(|00\rangle+|01\rangle),\ |\psi_2\rangle=|10\rangle,\ |\psi_3\rangle=|11\rangle\}$. And
four orthogonal states can be locally distinguished if and only if all of them are product states. There does not exist any set with more than four orthogonal states that can be distinguished under LOCC.
Therefore if a set of separable and orthogonal states $\{|\psi_i\rangle\}$ in two-qubit system is locally distinguishable, then the corresponding ensemble, $\rho_{s}=\sum_i p_i |\psi_i\rangle \langle \psi_i|$, $0\leq p_i\leq 1$, $\sum_i p_i=1$, is classical or semiquantum, which implies $Q_{12}(\rho_s)=0$ or $Q_{i}(\rho_s)=0$, $i=1,2$. The joint quantum correlation of $\rho_{s}$ is then zero, i.e. $\delta(\rho_{s})=0$. Namely, if the joint quantum correlation of an ensemble is positive, then these separable pure two-qubit states in the ensemble can not be locally distinguished.

\section{Summary}
To summarize, we have investigated the quantum correlations induced by the local von Neumann measurement and classified the quantum correlation into total quantum correlation, semiquantum correlation and joint quantum correlation. The properties of these quantum correlations have been discussed here. Furthermore an analytical formula of these quantum correlations for pure state has been obtained, which shows the quantum correlation in pure state is just the squared concurrence. Additionally, an experiment witness for these quantum correlations has been given in experimental accessible way. As applications, the nonexisting joint quantum correlation is shown to be the necessary condition for the local distinguishability of two-qubit separable and orthogonal pure states. This method can be generalized to multipartite case and we hope this work give insight into the comprehensive interpretation of the quantum correlation.

\end{document}